\documentclass[12pt,aps,pre,preprint,showpacs,showkeys]{revtex4}

\usepackage{graphicx}
\usepackage{subfigure}
\usepackage{amsmath}
\usepackage{amssymb}
\usepackage{bm}

\begin{document}

\title{XY model with higher-order exchange}
\author{Milan \v{Z}ukovi\v{c}}
 \email{milan.zukovic@upjs.sk}
\author{Georgii Kalagov}
 \affiliation{Institute of Physics, Faculty of Science, P. J. \v{S}af\'arik University, Park Angelinum 9, 041 54 Ko\v{s}ice, Slovakia}
\date{\today}

\begin{abstract}
An XY model, generalized by inclusion of up to an infinite number of higher-order pairwise interactions with an exponentially decreasing strength, is studied by spin-wave theory and Monte Carlo simulations. At low temperatures the model displays a quasi-long-range order phase characterized by an algebraically decaying correlation function with the exponent $\eta = T/[2 \pi J(p,\alpha)]$, nonlinearly dependent on the parameters $p$ and $\alpha$ that control the number of the higher-order terms and and the decay rate of their intensity, respectively. At higher temperatures the system shows a crossover from the continuous Berezinskii-Kosterlitz-Thouless to the first-order transition for the parameter values corresponding to a highly nonlinear shape of the potential well. The role of of topological excitations (vortices) in changing the nature of the transition is discussed.
\end{abstract}





\maketitle

\section{Introduction}

Mermin-Wagner theorem~\cite{merm66,hohe67} prevents any spontaneous breakdown of continuous symmetries for 2D systems with short-range interactions, such as a standard XY model. Nevertheless, it does not prevent a topological Berezinskii-Kosterlitz-Thouless (BKT) phase transition, due to the vortex-antivortex pairs unbinding~\cite{bere71,kost73}, to a quasi-long-range-order (QLRO) phase characterized by a power-law decaying correlation function.

Several modifications and generalizations of the XY model have been proposed, mostly by including higher-order terms to the Hamiltonian, motivated theoretically (critical properties and universality) as well as experimentally (modeling of some systems, such as liquid crystals~\cite{lee85,geng09}, superfluid A phase of $^3{\rm He}$~\cite{kors85}, and high-temperature cuprate superconductors~\cite{hlub08}). Inclusion of a biquadratic term, i.e., the system with the Hamiltonian ${\mathcal H}=-J_1\sum_{\langle i,j \rangle}\cos(\phi_{i,j})-J_2\sum_{\langle i,j \rangle}\cos(2\phi_{i,j})$, has been shown~\cite{lee85,kors85,carp89,shi11,hubs13,qi13} to lead to the separation of the dipole phase at lower and the quadrupole phase at higher temperature, for sufficiently large biquadratic coupling. The order-disorder phase transition was determined to belong to the BKT universality class, while the dipole-quadrupole phase transition had the Ising character. 

Recent series of studies~\cite{pode11,cano14,cano16} revealed that the model, in which the biquadratic term was generalized to a nematiclike coupling of the order $q>2$, i.e., ${\mathcal H}=-J_1\sum_{\langle i,j \rangle}\cos(\phi_{i,j})-(1-J_1)\sum_{\langle i,j \rangle}\cos(q\phi_{i,j})$ and $0 \leq J_1 \leq 1$, leads to a qualitatively different phase diagram for $q>3$, with additional ordered phases originating from the competition between the ferromagnetic and pseudonematic couplings and includes phase transitions belonging to the 2D Potts, Ising, or BKT universality classes.

Further generalization, motivated by orientational transitions in liquid crystals, lead to taking the $k$-th order Legendre polynomials of the dipole term, i.e., the Hamiltonian ${\mathcal H}=-\sum_{\langle i,j \rangle}P_k(\cos(\phi_{i,j}))$. With the increasing value of $k$, one may expect a qualitative change in the nature of the transition. In particular, a rigorous proof has been provided that the transition becomes first order for large enough values of $k$ in models with $O(n)$ symmetry for $n \geq 2$~\cite{ente02,ente05}. Nevertheless, for $O(2)$ case the studied values of $k=2$ and $4$ indicated that the behavior is always described by the BKT-like transition, just like in the standard XY model~\cite{fari05,berc05}. This is in contrast to the $O(3)$ case, in which a strong first-order phase transition was observed for $k=4$~\cite{mukh99,pal03}. 

Another non-linear model~\cite{doma84,himb84,blot02,sinh10a,sinh10b}, the potential shape of which can be controlled by a single parameter $p^2$, in the form ${\mathcal H}=2J\sum_{\langle i,j \rangle}(1-[\cos^2(\phi_{i,j}/2)]^{p^2})$, was introduced in effort to enable tuning its properties between the standard XY model belonging to the BKT universality and the $q$-state Potts model, which for large $q$ shows a first-order phase transition. Indeed, for large $p$ (proportional to the Potts $q$), such a model has been shown to undergo a first-order phase transition.

In the present study we introduce a generalized XY model that takes into account effects of up to an infinite number of higher-order (multipolar) terms with an exponentially vanishing influence. In spite of belonging to the same universality class (having same symmetry of the order parameter and same lattice dimensionality) as the standard XY model, we demonstrate that the model can display either the BKT or the first-order phase transition from the QLRO to the paramagnetic phase, depending on the parameters that control the degree of nonlinearity of the potential. 

\section{Model}

The considered model assumes only nearest-neighbor pairwise ferromagnetic interactions with the potential
\begin{equation}
\label{Potential}
H_{i,j}(p,\alpha)=-\sum_{k=1}^{p}J_{k}\cos^{k}\phi_{i,j},
\end{equation}
where $\phi_{i,j}=\phi_{i}-\phi_{j}$ is an angle between the nearest-neighbor spins and the respective exchange interactions decay as $J_k=\alpha^{-k}$, where $\alpha>1$. 

For an infinite number of the higher-order terms, i.e., $p \to \infty$, the Hamiltonian reduces to 
\begin{equation}
\label{Hamiltonian_inf}
{\mathcal H}(\alpha)=J(\alpha)\sum_{\langle i,j \rangle}H_{i,j}(\alpha)=-J(\alpha)\sum_{\langle i,j \rangle}\frac{\cos\phi_{i,j}}{\alpha-\cos\phi_{i,j}},
\end{equation}
where $\langle i,j \rangle$ denotes the sum over nearest-neighbor spins and $J(\alpha)=\alpha-1$ is an exchange interaction parameter chosen to normalize the weights $J_{k}$ (scaling them so they add up to 1).

For a finite number of the multipolar interaction terms, the system Hamiltonian can be expressed as
\begin{equation}
\label{Hamiltonian_fin}
{\mathcal H}(p,\alpha)=J(p,\alpha)\sum_{\langle i,j \rangle}H_{i,j}(p,\alpha)=-J(p,\alpha)\sum_{\langle i,j \rangle}\frac{\cos\phi_{i,j}\Big[1-\Big(\frac{\cos\phi_{i,j}}{\alpha}\Big)^{p}\Big]}{\alpha-\cos\phi_{i,j}},
\end{equation}
where $J(p,\alpha)=(\alpha-1)/(1-\alpha^{-p})$.

\begin{figure}[t!]
\centering
\subfigure{\includegraphics[scale=0.52,clip]{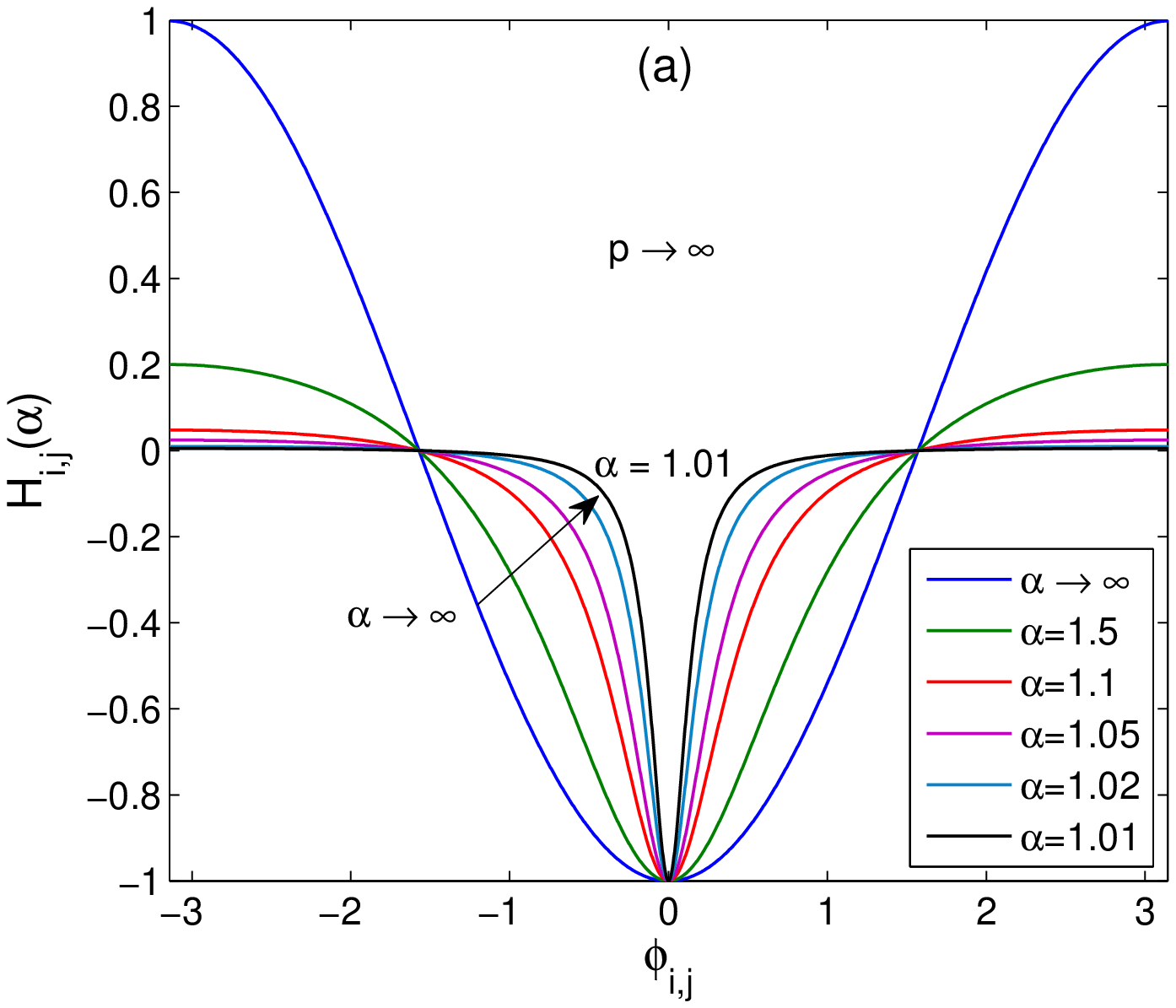}\label{fig:en_well_inf_a}}
\subfigure{\includegraphics[scale=0.52,clip]{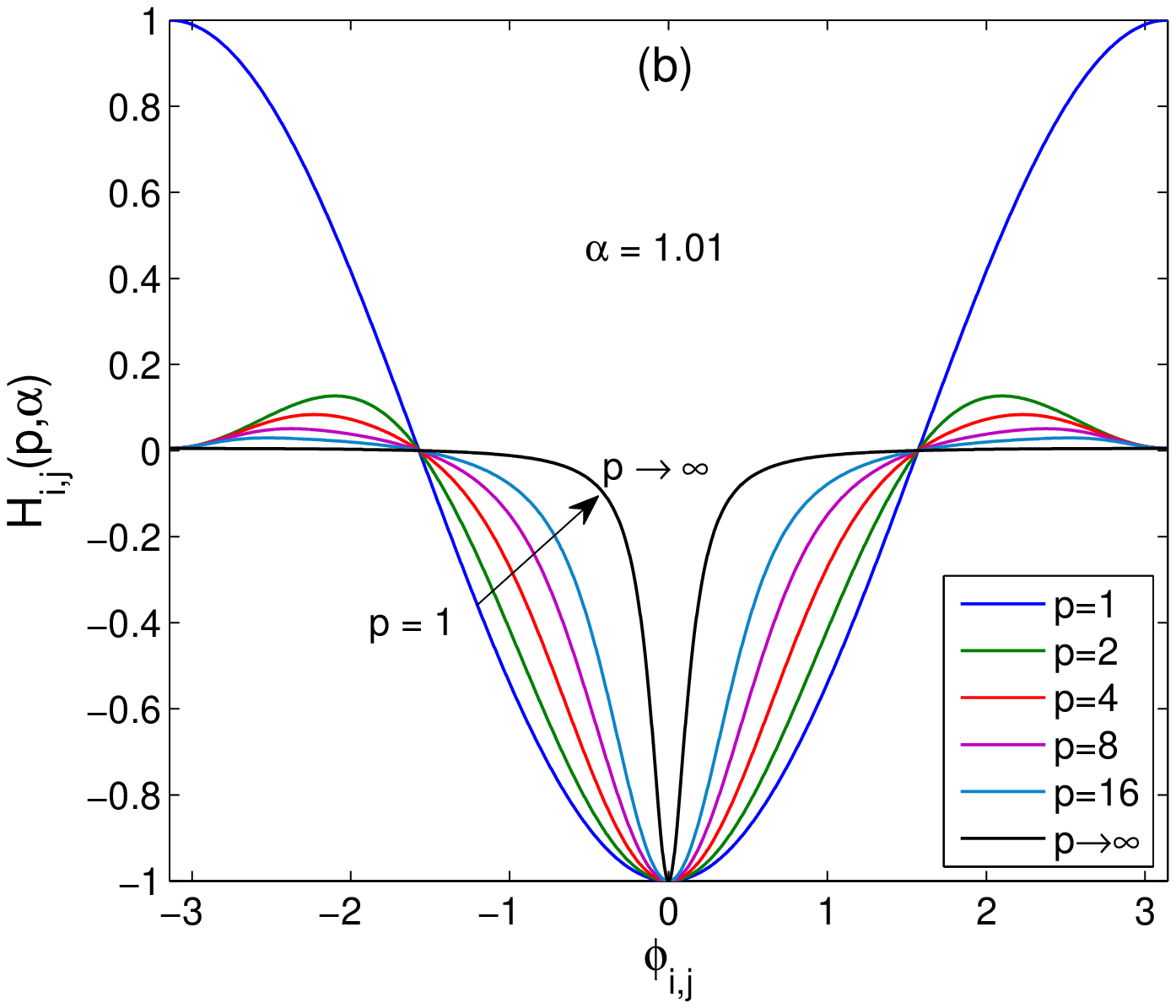}\label{fig:en_well_fin_a}}
\caption{(Color online) Potential functions of the cases of (a) $p \to \infty$ for several values of $\alpha$ and (b) a fixed $\alpha=1.01$ and various values of $p$.}\label{fig:well}
\end{figure}

Thus, while in the case of $p \to \infty$ there is only one parameter, $\alpha$, if the sum is truncated there are two parameters, $\alpha$ and $p$,  that can be used to change the shape of the respective potentials through changing the number of the higher-order terms and/or their weights. The shapes of the potentials in both cases are shown in Fig.~\ref{fig:well}, for different values of the parameters $\alpha$ and $p$. The case with $p \to \infty$ [Fig.~\ref{fig:en_well_inf_a}] reduces to the conventional XY model when the interaction terms decay extremely fast, i.e., for $\alpha \to \infty$, with the potential acquiring a cosine form. With the decrease in $\alpha$, the potential well gets narrower with a width tending to zero as $\alpha \to 1$. In the model with a finite $p$, a similar effect on the potential shape can be observed by increasing the number of the higher-order interaction terms, for sufficiently small values of $\alpha$ [Fig.~\ref{fig:en_well_fin_a}]. In this case, in the limit of $p \to \infty$ the width of the potential well will depend on the value of $\alpha$, as shown in Fig.~\ref{fig:en_well_inf_a}. 

It is worth noticing that for the case of a finite $p$ and a small $\alpha$, one can also observe a local minimum at $\phi=\pm\pi$ (see Fig.~\ref{fig:en_well_fin_a}). The latter is apparently related to the presence of the nematic term, the interaction strength of which is the second largest and for $\alpha \to 1$ it becomes comparable with the bilinear one. Therefore, care should be exercised when selecting the MC method particularly in the case of the presence of higher-order interactions with comparable strengths, when for $p>2$ even multiple local minima may develop, in order to prevent getting stuck in one of those especially at low temperatures.


\section{Methods}

\subsection{Spin wave approximation}
Let us consider a large scale asymptotic behavior of the two-point correlation function $g(x_1 - x_2) \equiv \langle\cos(\phi(x_1) - \phi(x_2) )\rangle =Re \langle \exp i \{\phi(x_1) - \phi(x_2)\} \rangle$ in the model defined through the more general form of the Hamiltonian, given by Eq.~(\ref{Hamiltonian_fin}).
Let $x$ be the coordinate vector of $i$-th spin, and $a$ be the lattice vector. At low temperatures one can assume smoothness of the field $\phi(x)$, and thus we may put $\phi(x+a) - \phi(x) = (a \cdot \nabla ) \phi(x) + \mathcal{O}( a^2)$. Having expanded Hamiltonian up to the second order in $a$, we find the low temperature approximation 
\begin{equation}
{\mathcal H}^{\rm sw} = J ^{\rm sw} \sum_{x} \sum_{a} \frac{1}{2} \{(a \cdot \nabla ) \phi(x)  \}^2 = J ^{\rm sw} \sum_{x} a^2 \frac{1}{2} \{ \nabla  \phi(x) \}^2 \rightarrow \frac{J ^{\rm sw}}{2} \int d^2 x \{\nabla \phi(x) \}^2,
\end{equation}
where $J ^{\rm sw} = \alpha/(\alpha-1) - p/(\alpha^p -1)$. We now see, that an asymptotic expression for the correlation function $g(x_1 - x_2)$ can be easily deduced by the Gaussian integration over all possible field configurations 
\begin{align}
g(x_1 - x_2) =  &\int \prod_x d \phi(x) \exp\left( -\frac{J ^{\rm sw}}{2} \int d^2 x \{\nabla \phi(x) \}^2  + i \{\phi(x_1) - \phi(x_2)\}  \right) =\\
=&\exp \left( -\frac{1}{J ^{\rm sw}} \int \frac{d^2 k}{(2 \pi)^2}\frac{1 - \exp( i k (x_1  - x_2) ) }{k^2} \right) = \exp\left(-\frac{1}{2 \pi J ^{\rm sw}} \ln \frac{e^{\gamma}|x_1 - x_2|}{2 a}  \right),
\end{align}
where $\gamma$ is the Euler-Mascheroni constant and the momentum integral has to be regularized in the ultra-violet region $0 \leq  |k| \lesssim 1/a$.        
As a result, large distance $|x_1-x_2| >> |a|$ power asymptotics reads as 
\begin{equation}
\label{cf_sw}\langle \cos (\phi(x_1) - \phi(x_2)) \rangle \sim \left( \frac{a}{|x_1 - x_2|}\right)^{\eta ^{\rm sw}},
\end{equation}
where the corresponding exponent $\eta ^{\rm sw} = T/(2 \pi J ^{\rm sw})$. We note that the resulting form of the correlation function exponent is also applicable to the specific case of the well studied bilinear-biquadratic model~\cite{lee85,kors85,carp89,shi11,hubs13,qi13}, with $p=2$ and $\alpha=J_1/J_2$.

\subsection{Monte Carlo}
We employ Monte Carlo (MC) simulations with the standard Metropolis dynamics for spin systems on a square lattice of a linear size $L$, imposing the periodic boundary conditions. For thermal averaging we take $N_{MC}$ MC sweeps after discarding another $N_0=0.2\times N_{MC}$ MC sweeps for thermalization. To obtain temperature dependencies of various thermodynamic quantities the simulations start in the paramagnetic phase at sufficiently high temperatures $T$ (measured in units $J/k_B$, where $k_B$ is the Boltzmann constant), and then proceed to lower temperatures with the step $\Delta T$. To maintain the system close to the equilibrium, at each $T-\Delta T$ simulations are initialized using the last configuration obtained at $T$.  

Close to the phase transition points we also perform finite-size scaling (FSS) analysis by using the reweighting techniques~\cite{ferr88,ferr89}, in order to identify the order and the universality class of the transition. Since in the criticality the integrated autocorrelation time $\tau$ is expected to dramatically increase, we make sure that sufficiently long simulation times are taken especially for larger lattice sizes. For reliable estimation of statistical errors we employed the $\Gamma$-method~\cite{wolf04}, that focuses on the explicit determination of the relevant autocorrelation functions and times, and gives more certain error estimates than for example the binning techniques. 

Typical values of the parameters are $L=24-72$, $N_{MC}=2 \times 10^5$ MC sweeps, and $\Delta T=0.025$, for the standard MC simulations, and up to $N_{MC}=10^7$ MC sweeps, for the reweighting. We avoided using larger lattice sizes, as tunneling times between the coexisting phases at first-order transitions can become enormous (see the inset of Fig.~\ref{fig:hist_alp_1_02a}).

We calculated the following quantities: the internal energy per spin $e=\langle {\mathcal H} \rangle/L^2$, 
the specific heat per site $c$
\begin{equation}
c=\frac{\langle {\mathcal H}^{2} \rangle - \langle {\mathcal H} \rangle^{2}}{L^2T^{2}},
\label{c}
\end{equation}
the magnetization
\begin{equation}
m=\langle M \rangle/L^2=\left\langle\Big|\sum_{j}\exp(i\phi_j)\Big|\right\rangle/L^2,
\label{m}
\end{equation}
the magnetic susceptibility
\begin{equation}
\label{chi}\chi = \frac{\langle M^{2} \rangle - \langle M \rangle^{2}}{L^2T}, 
\end{equation}
and the fourth-order magnetic Binder cumulant $U$
\begin{equation}
\label{U}U = 1-\frac{\langle M^{4}\rangle}{3\langle M^{2}\rangle^{2}}.
\end{equation} 
At the standard BKT to the paramagnetic phase transition the magnetization (susceptibility) is expected to vanish (diverge) as power law, characterized by the exponent $\eta=1/4$. The latter can be estimated by FSS of the respective quantities, as follows
\begin{equation}
\label{m_FSS}
m(L) \propto L^{-\eta/2},
\end{equation}
and
\begin{equation}
\label{xi_FSS}
\chi(L) \propto L^{2-\eta}.
\end{equation}
On the other hand, if the transition is of first order, then the internal energy $e$ and the magnetization $m$ will show a discontinuous behavior, the thermodynamic functions like the susceptibility $\chi$ are supposed to scale with volume, i.e., $\chi(L) \propto L^{2}$, and the Binder cumulant is expected to plunge to negative values~\cite{tsai98}.
 
A proper order parameter for the algebraic BKT phase is the helicity modulus $\Upsilon$ (or spin wave stiffness)~\cite{fish73,nels77,minn03}, which quantifies the resistance of the systems to a twist in the boundary conditions. It is defined as the second derivative of the free energy density of the system with respect to the twist $\tau$ along one boundary axis, which, for example, for the present XY model with the Hamiltonian~(\ref{Hamiltonian_inf}) results in the following expression
\begin{equation}
\label{helicity}
\Upsilon = \frac{1}{L^2}\sum_{\langle i,j \rangle_x} \frac{(\alpha-1) \alpha[2 \alpha \cos \phi_{i,j} + \cos (2\phi_{i,j}) - 3]}{2(\alpha-\cos \phi_{i,j})^3} - \frac{\beta}{L^2}\Big[\sum_{\langle i,j \rangle_x} \frac{(\alpha-1) \alpha \sin \phi_{i,j}}{(\alpha-\cos \phi_{i,j})^2}\Big]^2,
\end{equation}
where the summation $\sum_{\langle i,j \rangle_x}$ is taken over the nearest neighbors along the direction of the twist.

In order to directly study the topological excitations (defects) we evaluate a defect density $\rho$. Let us recall that a vortex (antivortex) is a topological defect which corresponds to the spin angle change by $2\pi$ $(-2\pi)$ going around a closed contour enclosing the excitation core. In the MC simulation they are identified by summation of the angles between adjacent four spins on each square plaquette for each equilibrium configuration. Thus, the summation equal to $2\pi$, $-2\pi$ and $0$ means that in the plaquette there is a vortex, antivortex and no topological defect, respectively~\footnote{We allow for a small deviation from these values due to numerical errors.}. Then the defect density $\rho$ is obtained as a thermodynamic average of the absolute value of the vorticity (taking into consideration both vortices and antivortices) summed over the entire lattice and normalized by the system volume $L^2$.

\section{Low-temperature behavior}

\begin{figure}[t!]
\centering
\subfigure{\includegraphics[scale=0.52,clip]{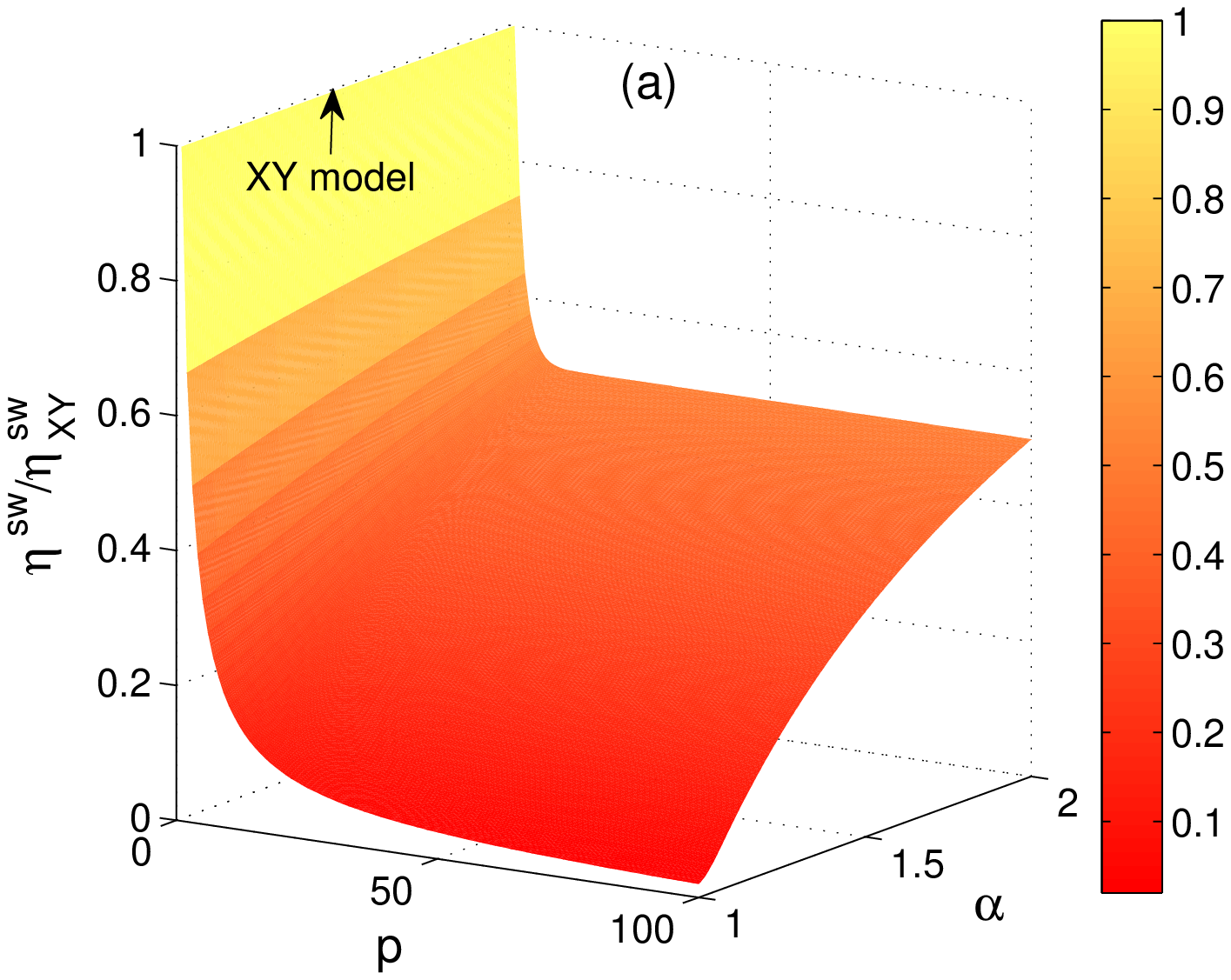}\label{fig:eta-p-alp}}
\subfigure{\includegraphics[scale=0.52,clip]{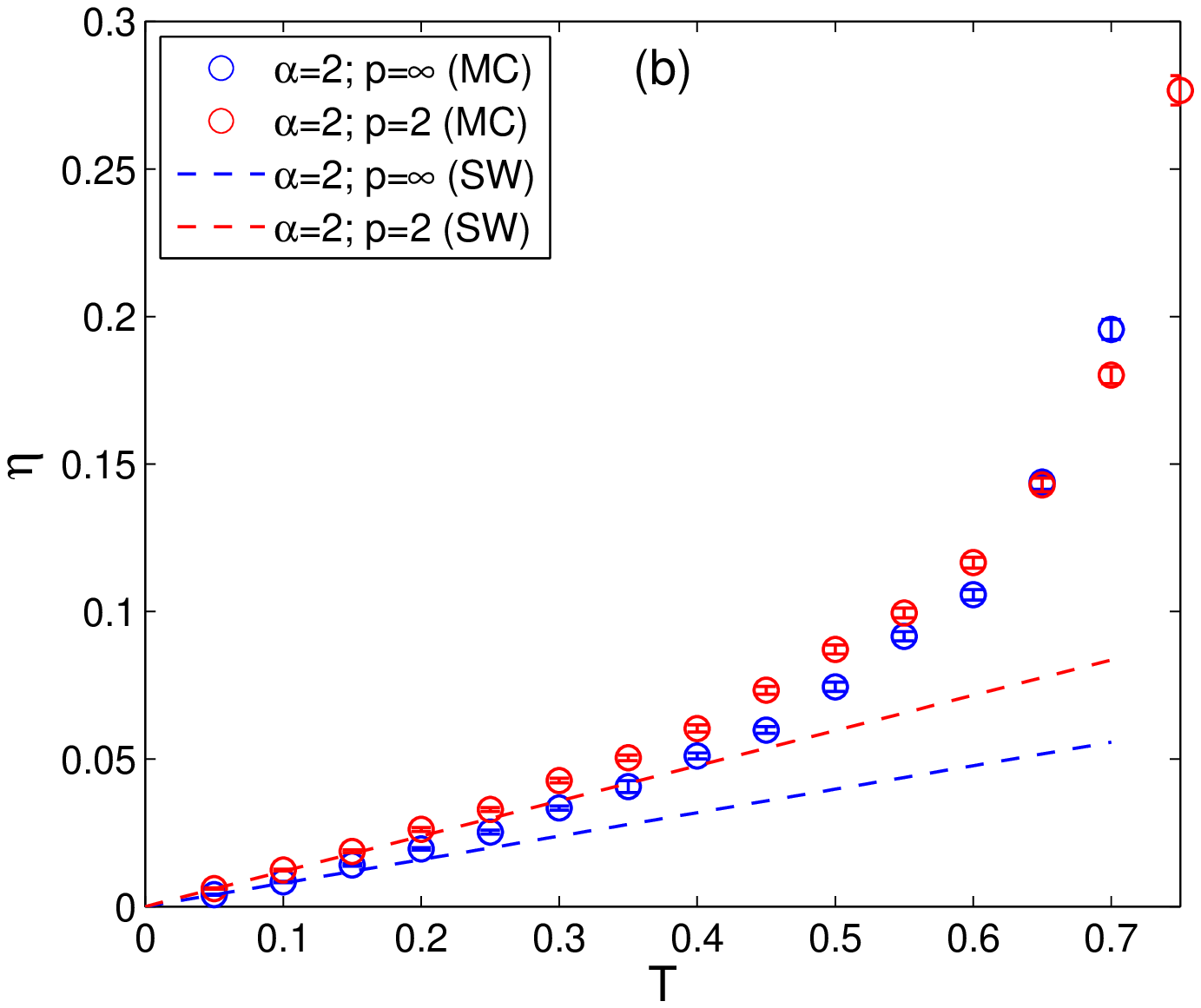}\label{fig:eta-T}}
\caption{(Color online) (a) SW approximation of the correlation function exponent $\eta^{\rm sw}$ normalized per that of the XY model, shown in the $(p-\alpha)$ parameter plane. (b) The exponent $\eta$ as a function of temperature, obtained from the SW theory (dashed lines) and MC simulations (symbols), for selected parameter values.}\label{fig:low-T}
\end{figure} 

The spin-wave approximation predicts the existence of the QLRO phase characterized by a power-law decaying correlation function, given by Eq.~\ref{cf_sw}. The exponent $\eta^{\rm sw}$ is formally similar to that of the standard XY model $\eta^{\rm sw}_{\rm XY}$, i.e., linearly dependent on the temperature, however, through the interaction $J^{\rm sw}$ it is also nonlinearly dependent on the parameters $p$ and $\alpha$. The reduced exponent $\eta^{\rm sw}/\eta^{\rm sw}_{\rm XY}=J^{\rm sw}_{\rm XY}/J^{\rm sw}$ as a function of the parameters $p$ and $\alpha$ is depicted in Fig.~\ref{fig:eta-p-alp}. One can notice that inclusion of just a few higher-order interaction terms causes a drastic drop of the exponent, followed by a leveling off if their couplings relative to the bilinear term are very small, i.e., for larger $\alpha$. On the other hand, if the interactions at the higher-order terms are comparable with the bilinear one, i.e., for $\alpha \to 1$, the exponent is further decreased with inclusion of more and more terms.

We also confront the spin-wave theory exponents $\eta^{\rm sw}$ with those obtained from MC simulations, for selected parameter values. In Fig.~\ref{fig:eta-T} we show temperature dependencies of both $\eta^{\rm sw}$ and $\eta^{\rm mc}$, for two cases of $(\alpha,p)=(2,2)$ and $(\alpha,p)=(2,\infty)$. As expected, the correspondence is very good at low temperatures but for $T \gtrsim 0.15$ the spin-wave approximation apparently underestimates the exponent values.

\section{Phase transitions}
\subsection{Infinite series model}

\begin{figure}[t!]
\centering
\subfigure{\includegraphics[scale=0.52,clip]{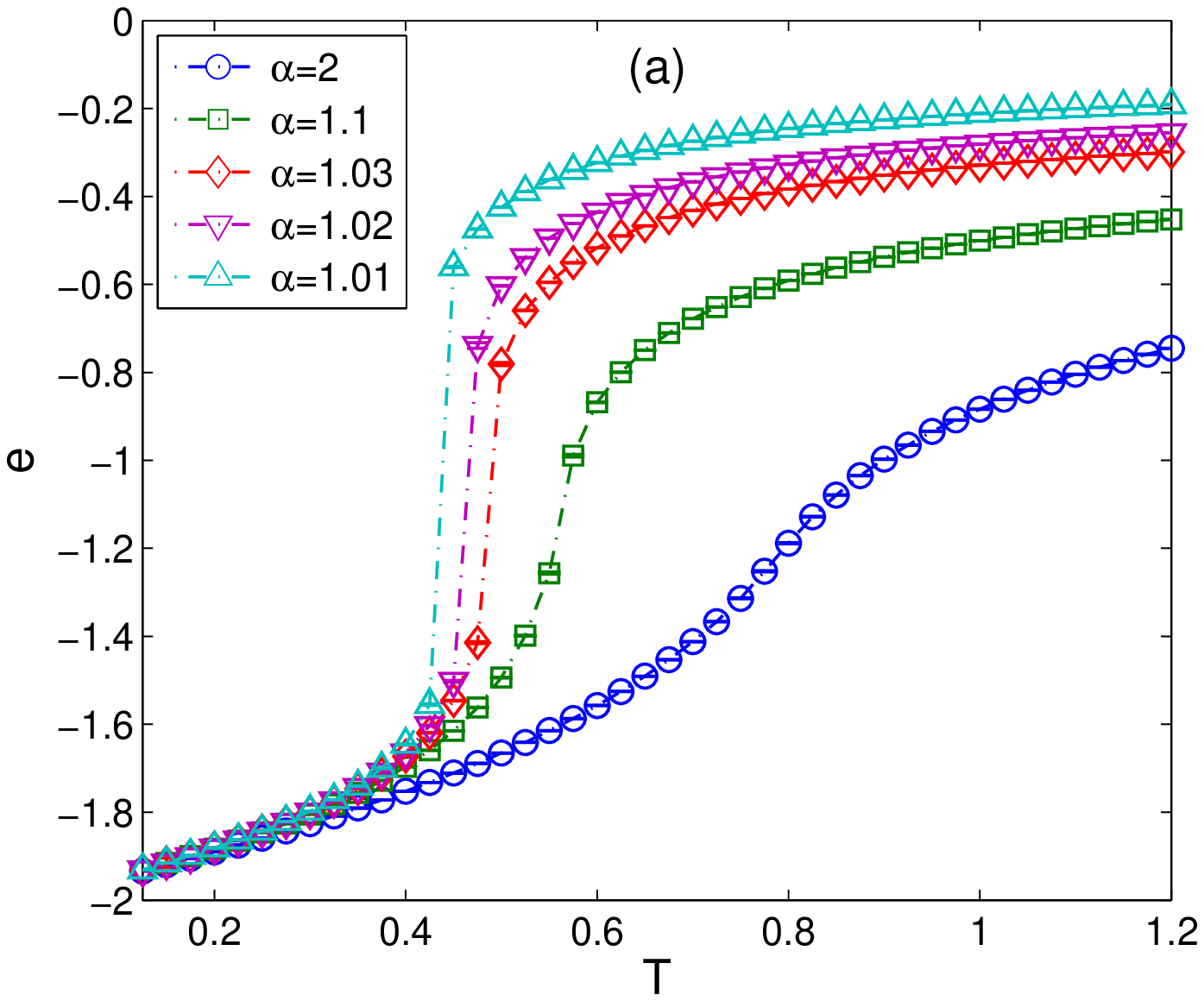}\label{fig:e-T}}
\subfigure{\includegraphics[scale=0.52,clip]{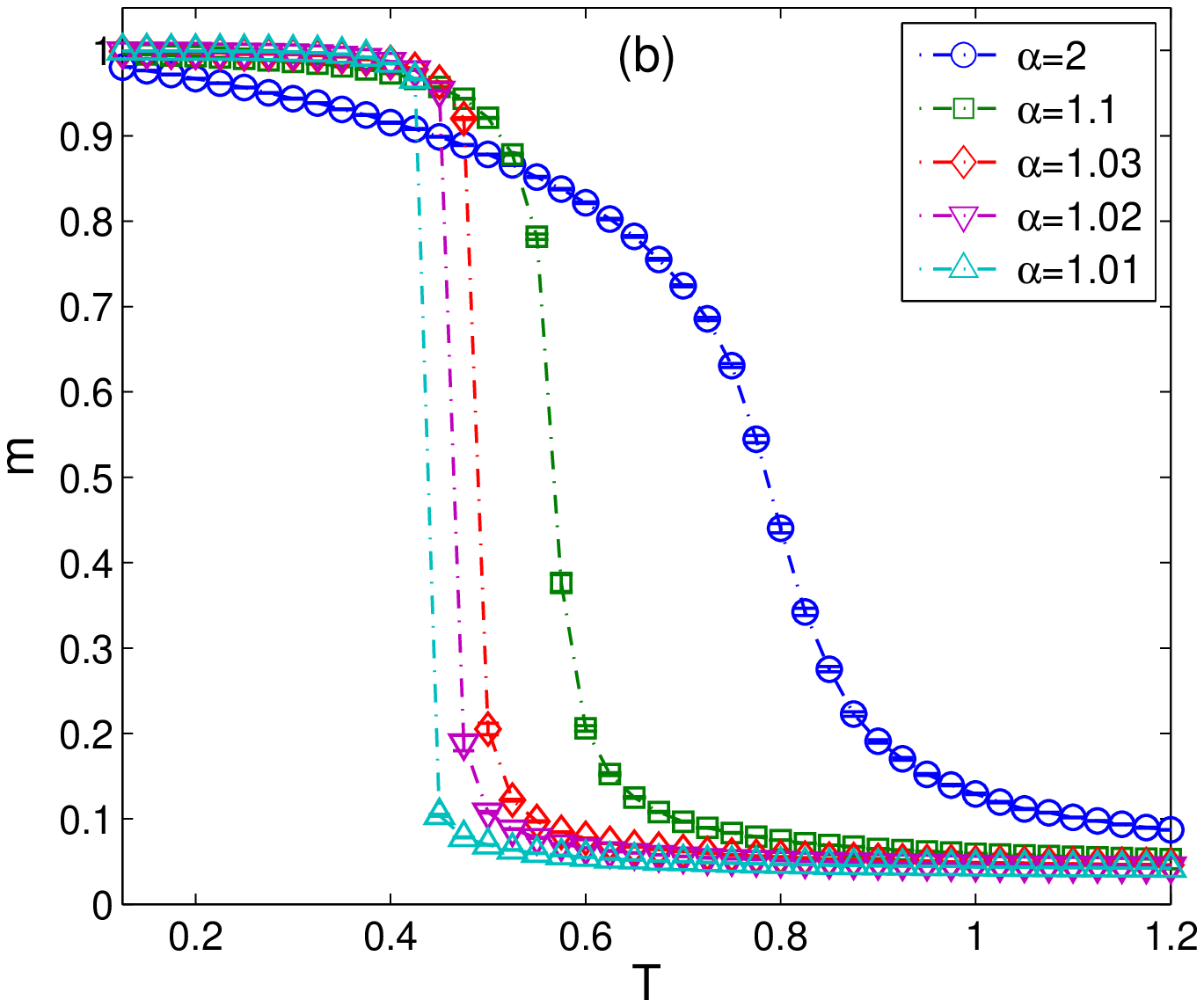}\label{fig:m-T}}
\subfigure{\includegraphics[scale=0.52,clip]{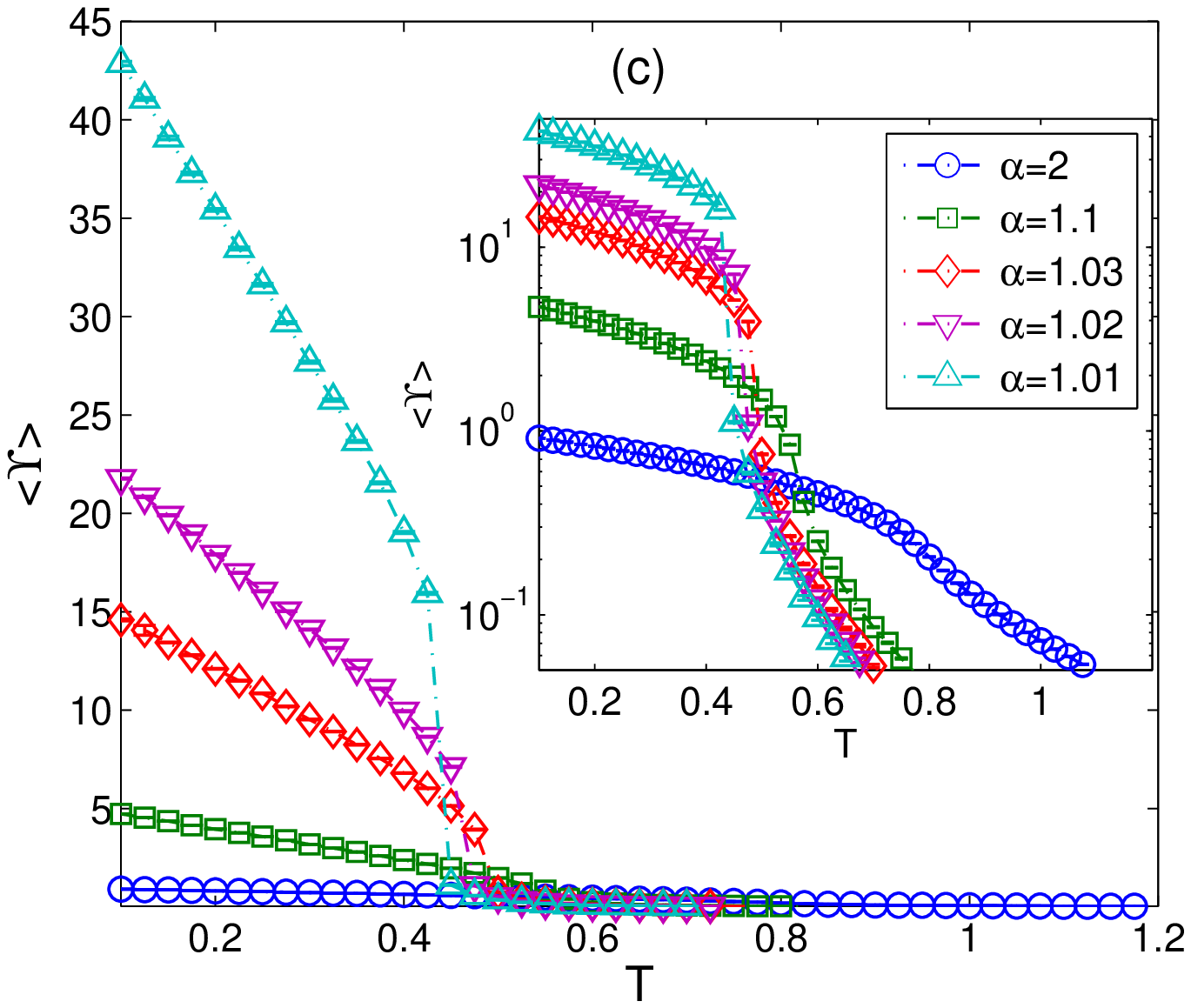}\label{fig:hm-T}}
\caption{(Color online) Temperature dependencies of the internal energy, magnetization and helicity modulus, for $L=24$, $p \to \infty$, and several values of $\alpha$. In (c), the inset shows the same figure on a log-log scale.}\label{fig:x-T}
\end{figure} 

The effect of a varying parameter $\alpha$ on magnetic and thermodynamic properties of the model can be observed in Fig.~\ref{fig:x-T}, in which temperature dependencies of the internal energy, the magnetization and the helicity modulus are plotted for various values of $\alpha$ and a fixed value of $L=24$. For $\alpha=2$, the effect of the higher-order terms in the Hamiltonian is almost negligible and the behavior of all the quantities resembles that of the standard XY model. Namely, they show a smooth variation in the vicinity of the transition point, as expected for the BKT transition. With decreasing $\alpha$ the effect of the higher-order terms becomes more pronounced and makes changes of the quantities at the transition more dramatic. In particular, as $\alpha$ approaches the limiting value of one, all start showing an apparently discontinuous behavior, typical for a first-order phase transition.

\begin{figure}[t!]
\centering
\subfigure{\includegraphics[scale=0.52,clip]{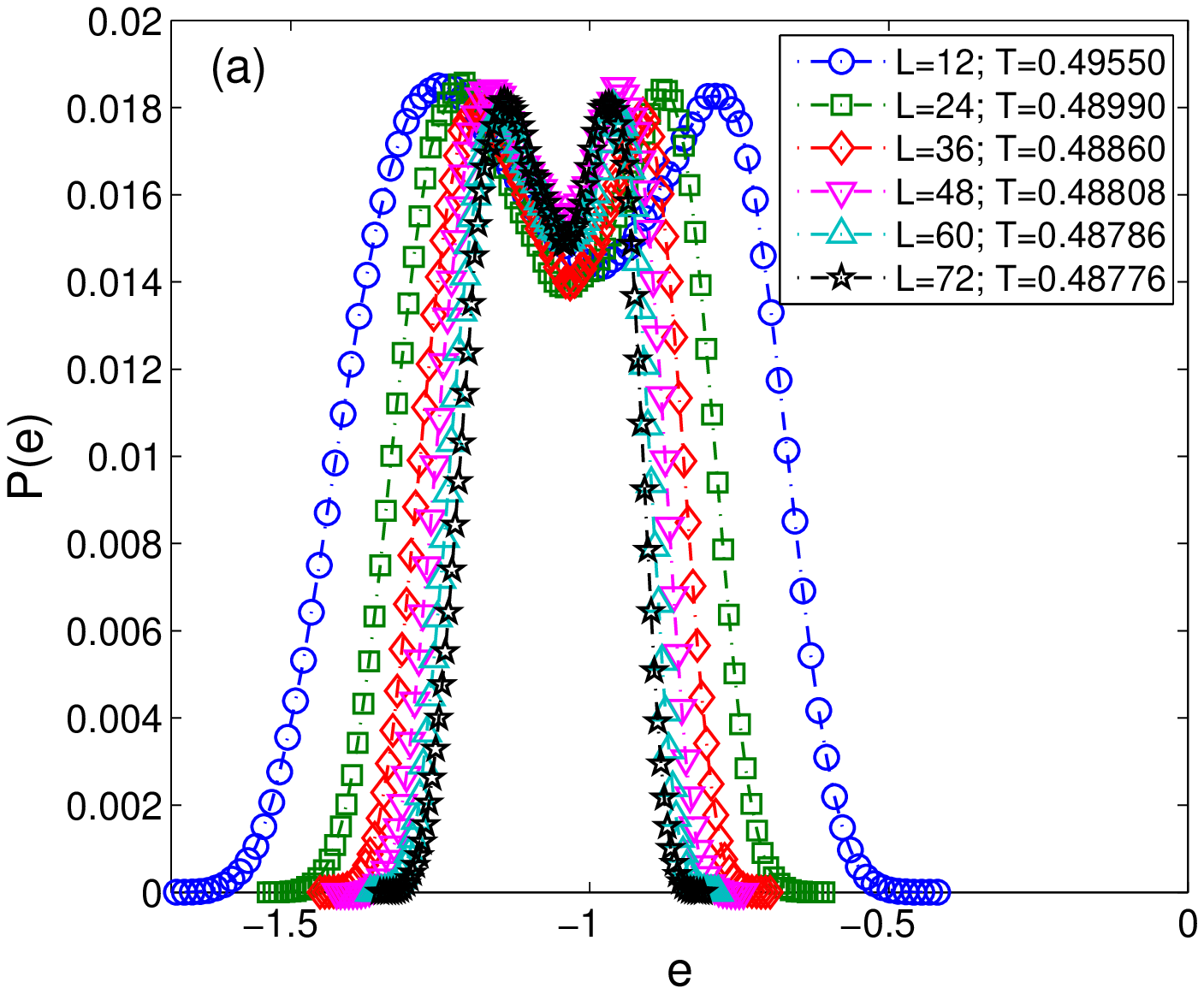}\label{fig:hist_alp_1_03a}}
\subfigure{\includegraphics[scale=0.52,clip]{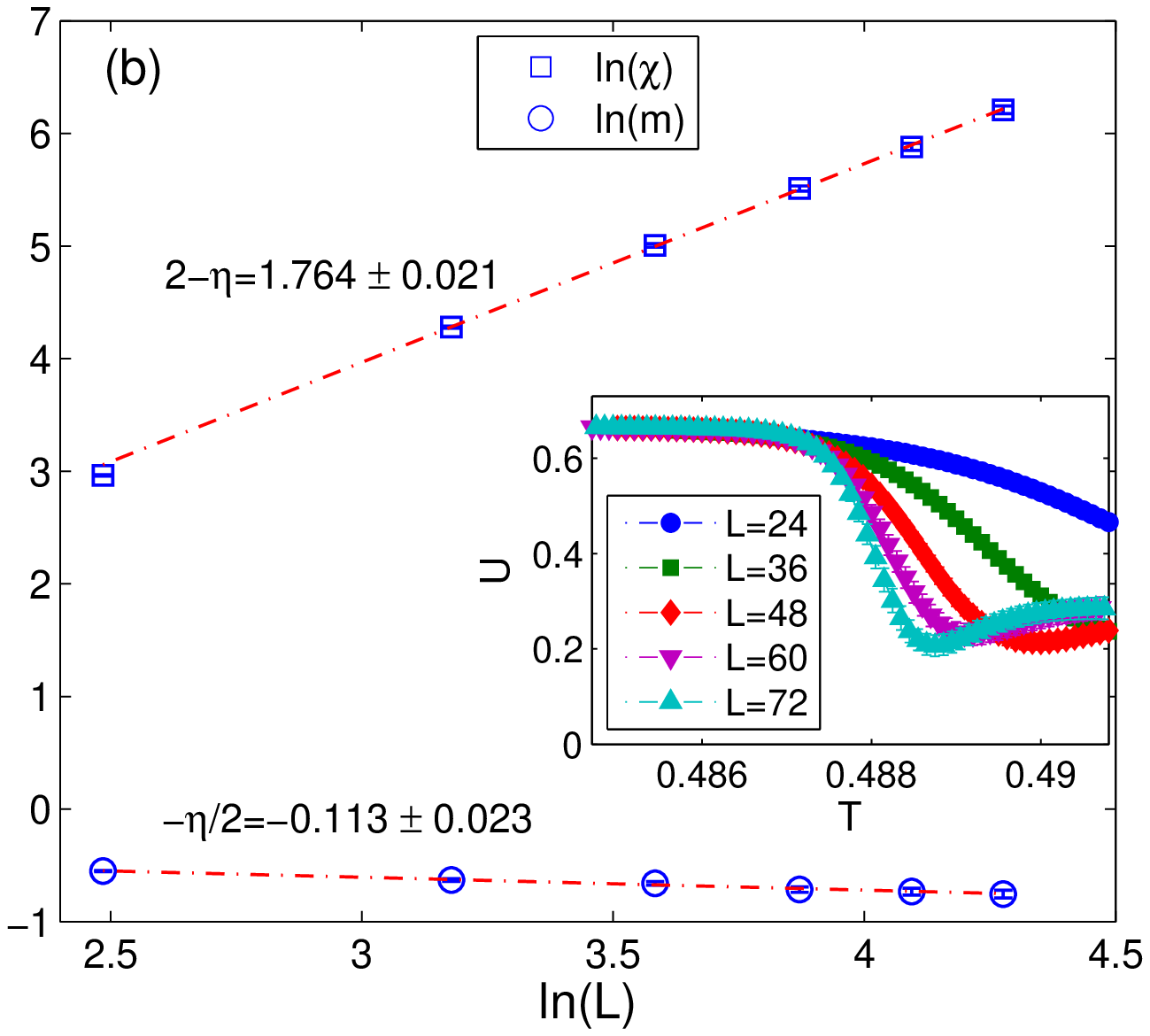}\label{fig:fss_chi_alp1_03}}\\
\subfigure{\includegraphics[scale=0.52,clip]{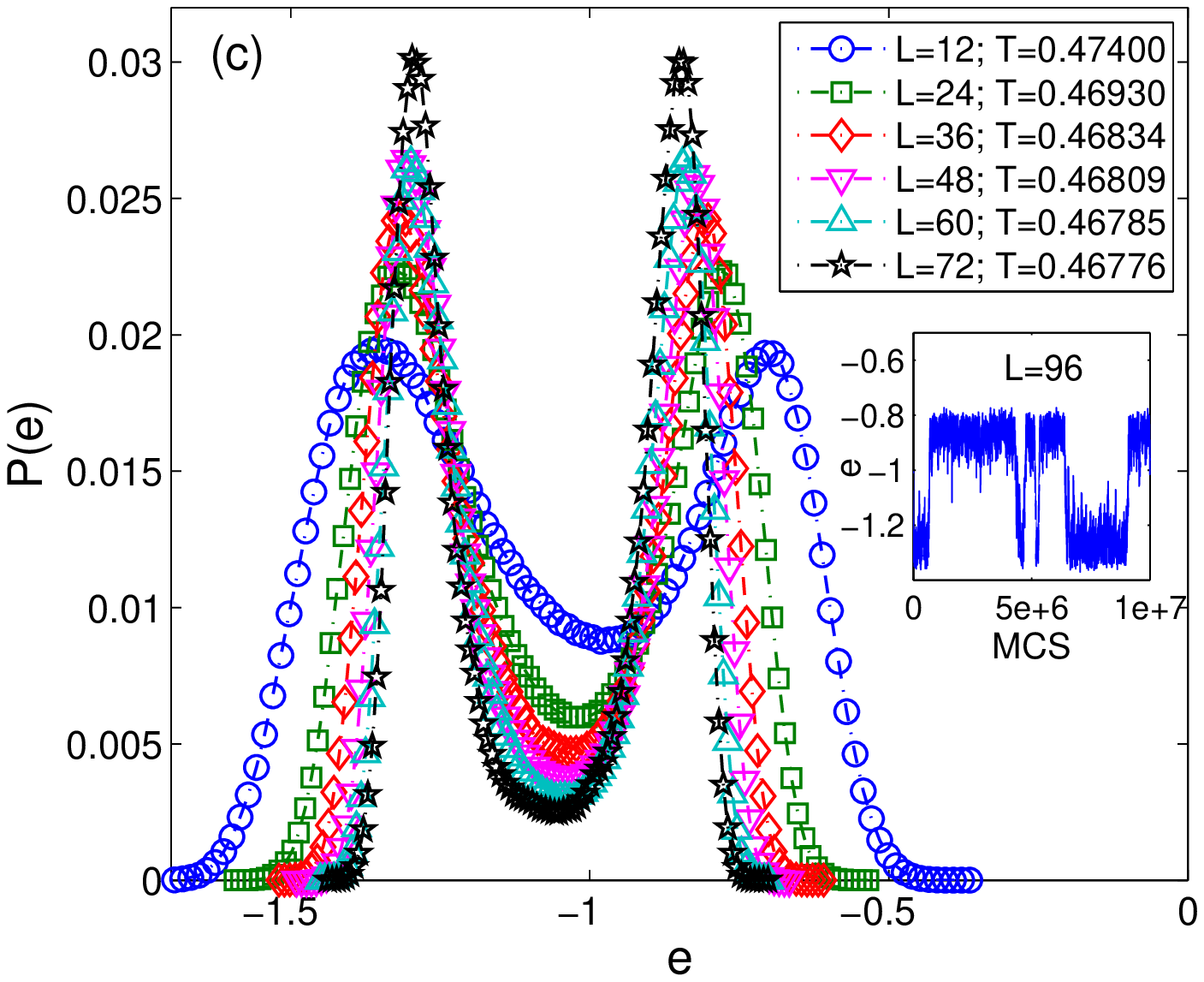}\label{fig:hist_alp_1_02a}}
\subfigure{\includegraphics[scale=0.52,clip]{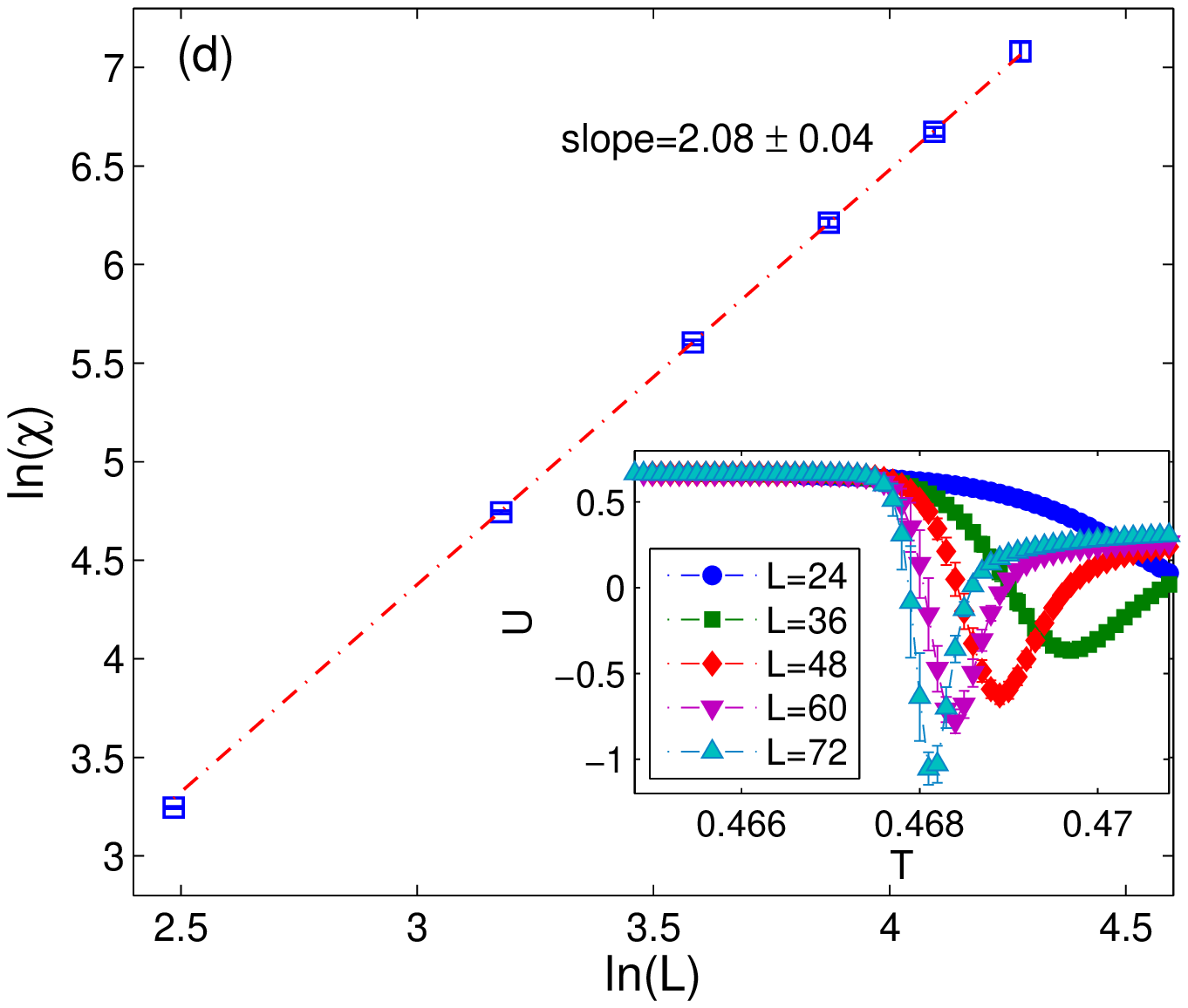}\label{fig:fss_chi_alp1_02}}
\caption{(Color online) Energy histograms and FSS analysis for (a,b) $\alpha=1.03$ and (c,d) $\alpha=1.02$. The histograms are reweighted to the temperatures at which the peaks are of equal height. The insets in the right panels show the respective Binder cumulants. The inset in panel (c) demonstrates huge tunneling times for larger sizes, e. g., for $L=96$ they are of the order of $10^6$ MCS.}\label{fig:hist_FSS_inf}
\end{figure}

In order to confirm that the observed behavior indeed corresponds to the crossover from the continuous to the first-order transition, next we study the character of the energy distribution and perform a FSS analysis in the concerned region of the parameter space. In Fig.~\ref{fig:hist_FSS_inf} we present the results for $\alpha=1.03$ (a,b) and $\alpha=1.02$ (c,d). In the left panels, the plots of the energy histograms for different sizes $L$ are reweighted to the temperature at which both peaks are of equal height. In both cases, the plots indicate a bimodal distribution that is characteristic for a discontinuous first-order transition. Nevertheless, there is a significant difference between them. We note that at the first-order transition as $L$ increases the heights of the peaks are expected to increase at the cost of the dip (barrier) between them, that should tend to zero and the distance between the peaks should approach a finite value, corresponding to the latent heat released at the discontinuous transition. This is exactly what we witness in the case of $\alpha=1.02$ [Fig.~\ref{fig:hist_alp_1_02a}], however, the behavior for $\alpha=1.03$ is quite different. Namely, from Fig.~\ref{fig:hist_alp_1_03a} we can see that with the increasing lattice size the height of the peaks virtually does not change, the dip between them does not get deeper and it becomes narrower as the peaks continue to move towards each other. Thus we believe that the observed double-peak structure for $\alpha=1.03$ is just a finite-size effect and in the thermodynamic limit it will vanish. We note that such a pseudo-first-order behavior was also observed in some other systems, such as the 4-state Potts and $J_1-J_2$ Ising models~\cite{jin12}.

The above conjecture is furthermore corroborated by FSS analysis and the behavior of the Binder cumulant. In particular, for the case of $\alpha=1.03$ the FSS relations [Eqs.~\ref{m_FSS} and~\ref{xi_FSS}] give the estimate of the exponent $\eta$ in accordance with the value $1/4$ expected for a standard BKT phase transition [Fig.~\ref{fig:fss_chi_alp1_03}], while for $\alpha=1.02$ the magnetic susceptibility scales with volume [Fig.\ref{fig:fss_chi_alp1_02}], as it should be in the case of a first-order transition. A smooth variation of the Binder cumulant within positive values in the former case and an abrupt descent to negative values in the latter case (see insets) provide additional evidence for such a scenario.  

\begin{figure}[t!]
\centering
\subfigure{\includegraphics[scale=0.52,clip]{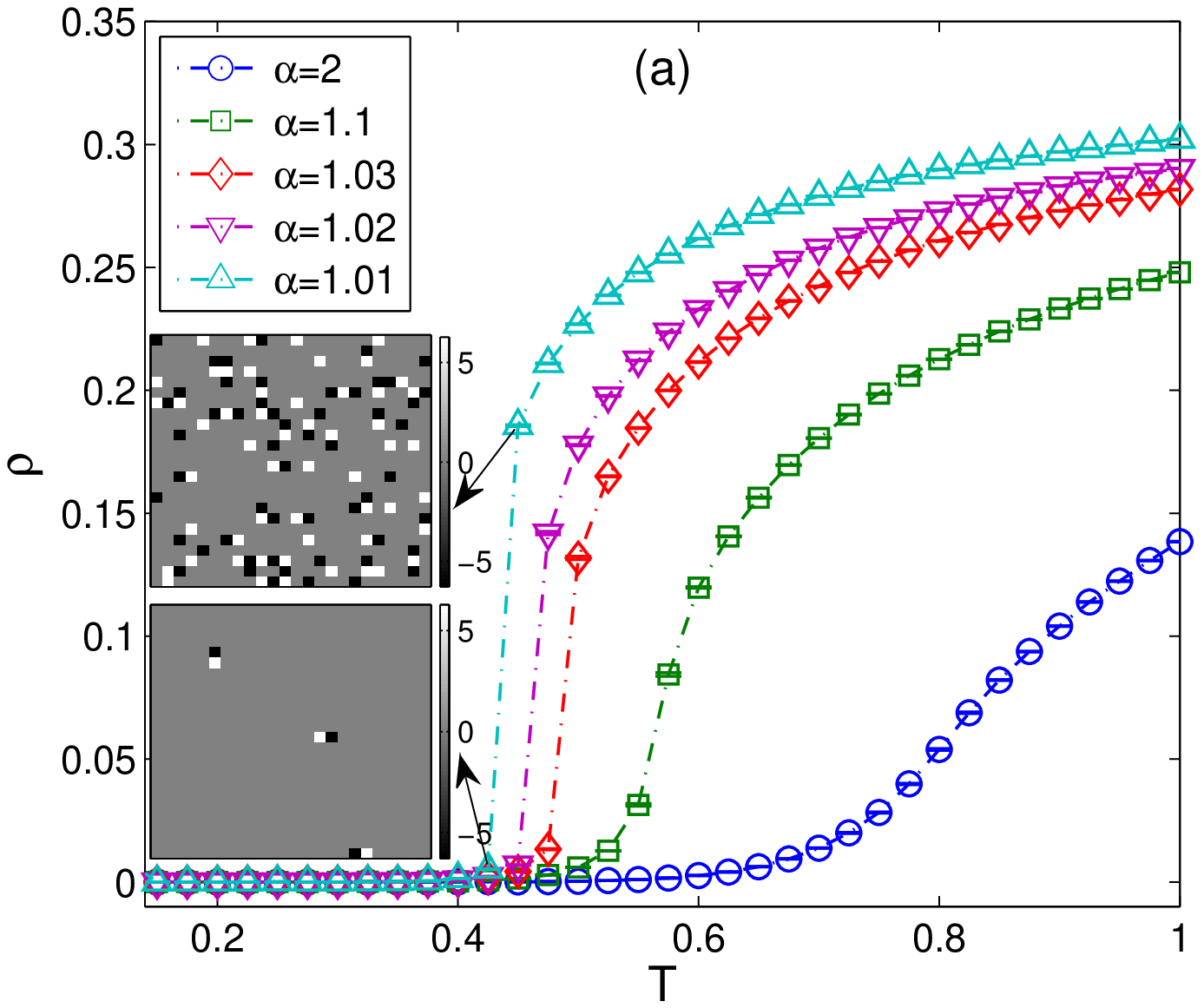}\label{fig:rho-T}}
\subfigure{\includegraphics[scale=0.52,clip]{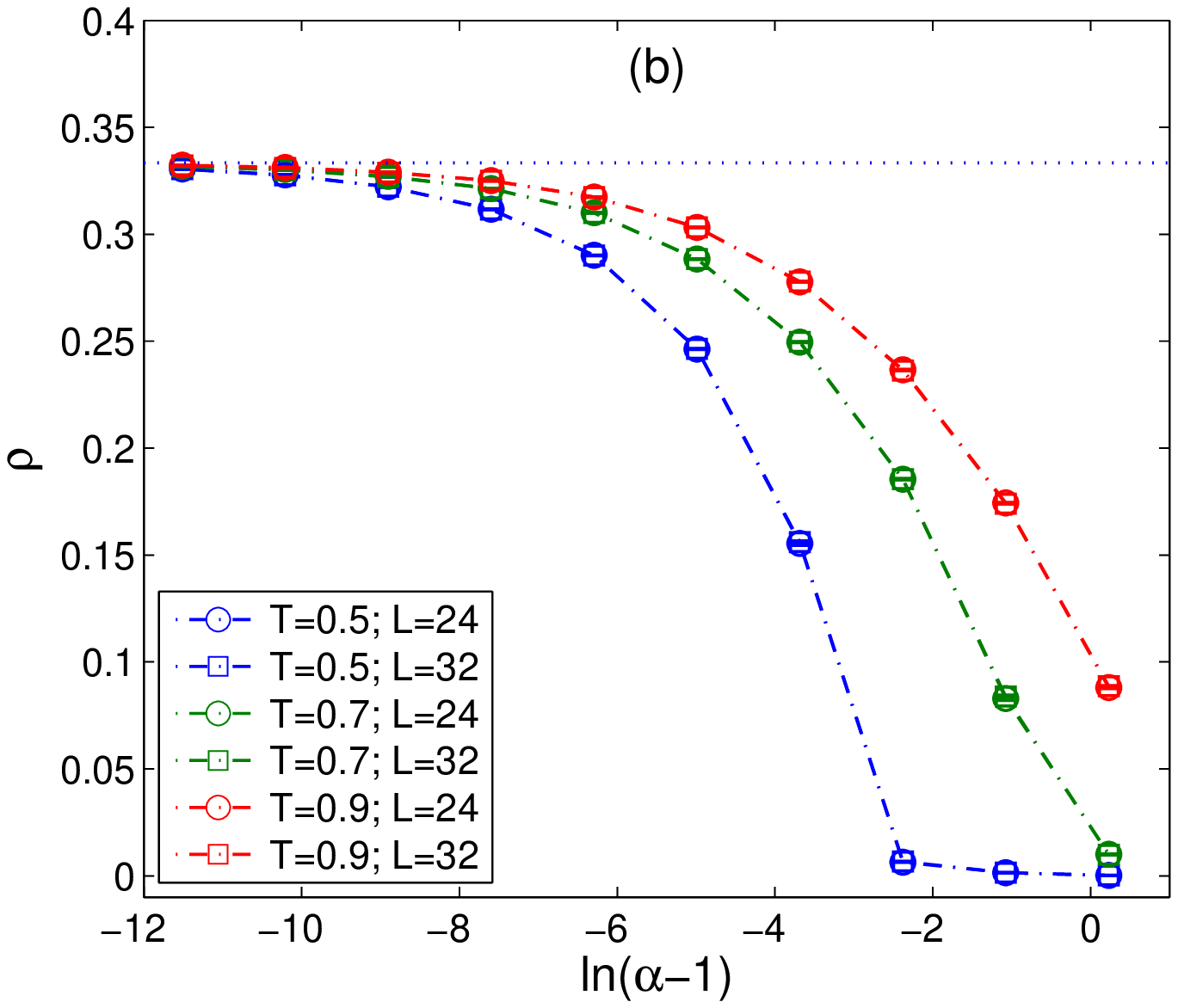}\label{fig:rho-alp}}
\caption{(Color online) (a) The defect density $\rho$ as a function of temperature, for several values of $\alpha$. The insets show typical snapshots just below (lower panel) and just above (upper panel) the transition point, depicting vortices (white squares) and antivortices (black squares), for $\alpha=1.01$. (b) $\rho$ as a function of $\alpha$, for three values of $T$ and two values of $L$.}\label{fig:rho-T-alp}
\end{figure}

The crossover to the first-order behavior can be understood by elucidation of the role of the topological defects in a varying potential shape, tuned by the parameter $\alpha$. In Fig.~\ref{fig:rho-T} we present temperature dependencies of the defect density $\rho$, for selected values of $\alpha$. It is evident that at the transition temperature from the BKT to the paramagnetic phase $\rho$ anomalously increases. The increase becomes particularly dramatic (resembling a jump) for the values of $\alpha$ close to one. A sudden increase of the defects at the transition for $\alpha=1.01$ is illustrated in the insets of Fig.~\ref{fig:rho-T}. The snapshot in the lower panel is taken just below the transition temperature and shows just a few vortex-antivortex pairs. The snapshot in the upper panel, taken just above the transition point, shows a great number of dissociated vortices (white squares) and antivortices (black squares). 

It is also interesting to study the behavior of topological excitations with the parameter $\alpha$. In Fig.~\ref{fig:rho-alp} we show dependences of the defect density $\rho$ on $\alpha$, for selected temperatures $T=0.5,0.7$ and $0.9$. One can notice a sharp increase of the defect density as $\alpha \to 1$ (note the semi-logarithmic scale), which seems to approach a common saturation value of $\rho_s=1/3$ (dotted line). Two sets of curves obtained for two different $L=24$ and $32$ that almost collapse on each other demonstrate that the behavior is practically independent of the lattice size.

Similar behavior has also been reported for the modified XY model, introduced by Domany {\it et al.}~\cite{doma84}, and explained in the later studies~\cite{himb84,sinh10b}. The abrupt increase of the defects, resulting in a first-order transition, is related to the shape of the potential well. Namely, for certain values of the parameter the well becomes very narrow which suppresses formation of defect pairs at low temperatures and thus facilitates their dramatic proliferation at the transition point. We believe that similar mechanism is responsible for the crossover to the first-order transition also in the present model. The nonlinearity of the potential well is controlled by the parameter $\alpha$ and, as shown in Fig.~\ref{fig:en_well_inf_a}, for the values close to one it becomes narrow enough to lead to the discontinuous phase transition. 

We note that besides the integer vortices studied above, it is reasonable to assume also the presence of various fractional vortices, resulting from the higher-order terms. Since our model involves a large number of them we did not attempt to evaluate all their individual densities. Nevertheless, in Fig.~\ref{fig:hm-T} one can see that in the temperature dependencies of the helicity modulus there are no anomalies, such as, for example, in Ref.~\cite{cano16}, except the one related to the transition to the paramagnetic state. This fact along with the behavior of other evaluated quantities, indicates that the integer and fractional vortices unbind at the same temperature corresponding to the transition point between the BKT and paramagnetic phases.

\begin{figure}[t!]
\centering
\includegraphics[scale=0.52,clip]{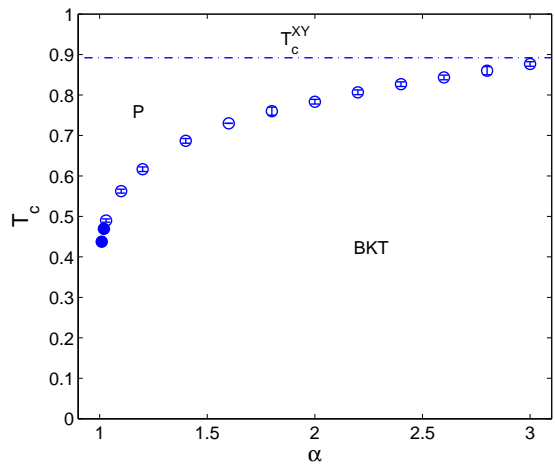}
\caption{(Color online) Phase boundary as a function of the parameter $\alpha$, separating the BKT and paramagnetic (P) phases. The (pseudo)transition temperatures are obtained from maxima of the specific heat curves, for $L=24$. The filled symbols represent the first-order transition points and the dashed line the transition temperature of the standard XY model.}\label{fig:PD}
\end{figure} 

Finally, the approximate phase diagram in $T-\alpha$ parameter plane is depicted in Fig.~\ref{fig:PD}. Rough estimates of (pseudo)transition temperatures are obtained as positions of maxima of the specific heat curves from several independent MC runs, for $L=24$~\footnote{Similar values could be obtained by considering positions of maxima of the magnetic susceptibility instead of the specific heat.}. The filled circles represent the first-order transition points at $\alpha=1.01$ and $1.02$, and the dashed line shows the transition temperature of the standard XY model, which is expected to be recovered in the limit of $\alpha \to \infty$. We note that these pseudo-transition temperatures slightly overestimate the true thermodynamic limit values [see, e.g., Figs.~\ref{fig:hist_alp_1_03a} and~\ref{fig:hist_alp_1_02a}]. Overall, the decreasing $\alpha$ shifts the transition temperature from the paramagnetic (P) to the BKT phase to lower values and eventually also changes the nature of the transition to the first-order one.

\subsection{Truncated series model}
Above we demonstrated that the first-order transition is a result of the increased influence of higher-order terms. Next, we will be interested in whether their infinite number is an indispensable ingredient for the first-order character of the transition or it can also persist when only a finite number of the terms is considered. We showed that for $p \to \infty$ the first-order transition exists if $\alpha \gtrsim 1$. On the other hand, the case of $p=2$ is well know to show the standard BKT transition for any value of $\alpha$~\cite{lee85}. Therefore, for a fixed $\alpha \gtrsim 1$ one can expect a crossover between the two regimes at some value of $p_c$.
 
\begin{figure}[t!]
\centering
\subfigure{\includegraphics[scale=0.52,clip]{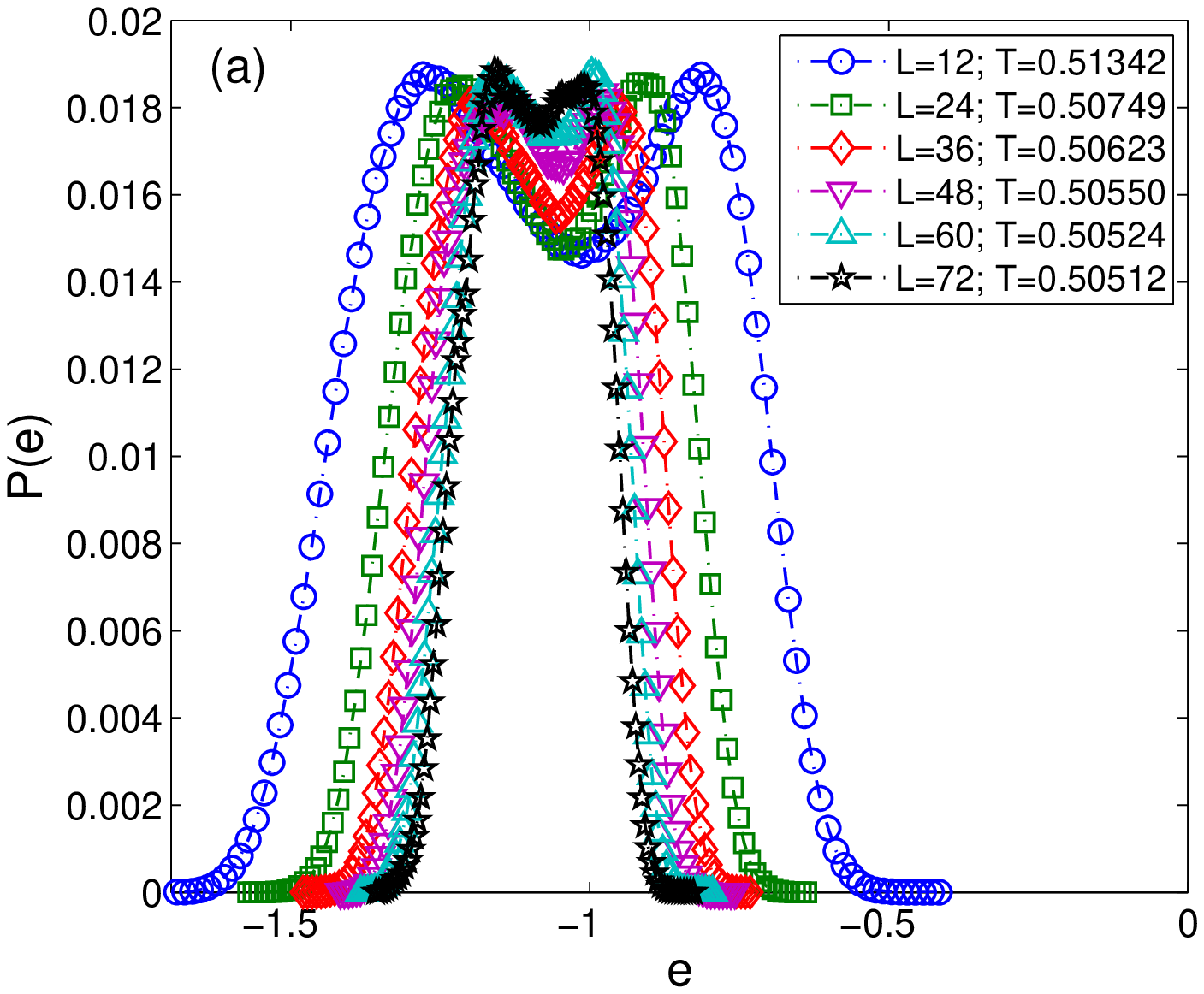}\label{fig:hist_alp_1_01_m50a}}
\subfigure{\includegraphics[scale=0.52,clip]{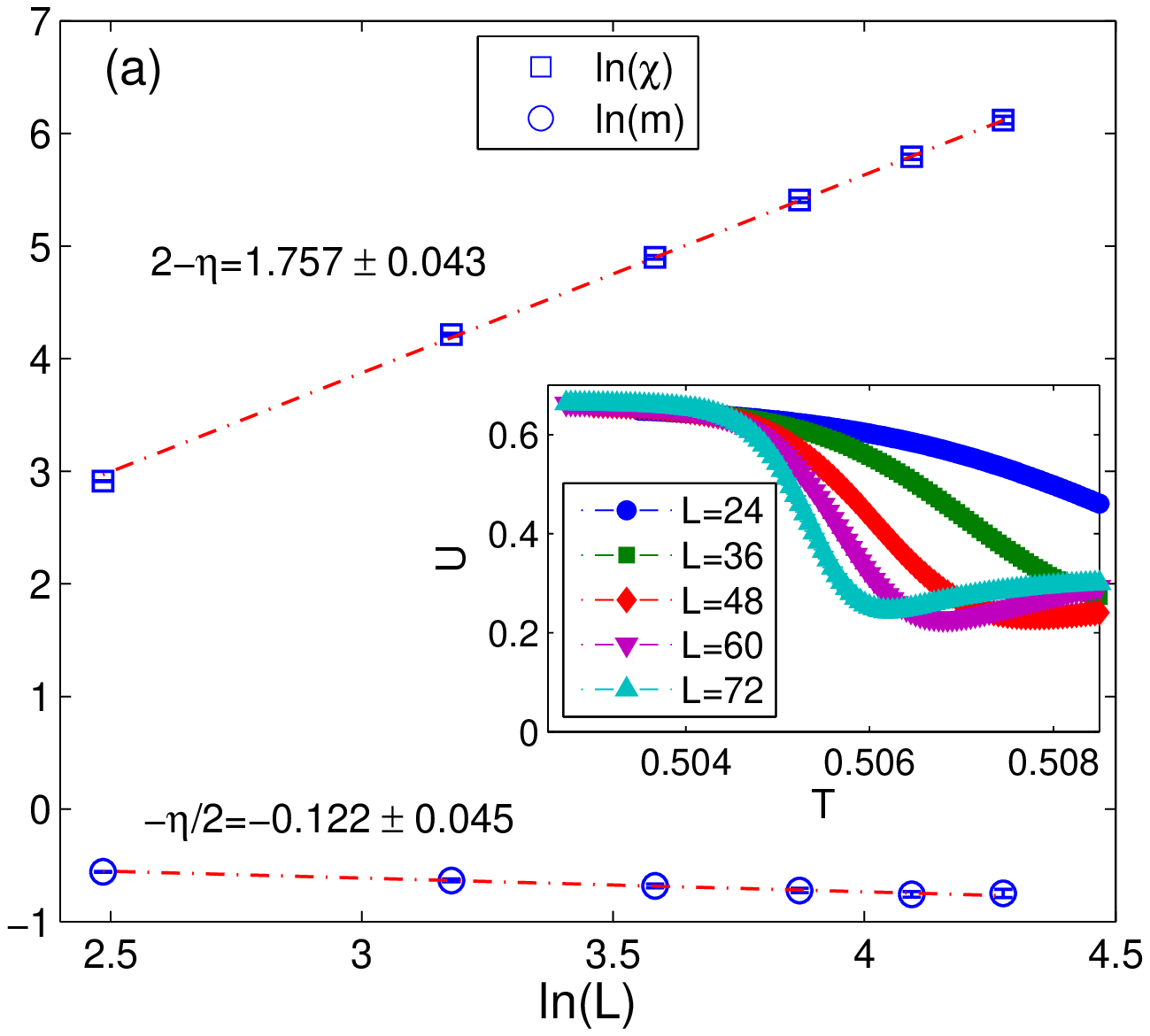}\label{fig:fss_chi_alp1_01_m50}}\\
\subfigure{\includegraphics[scale=0.52,clip]{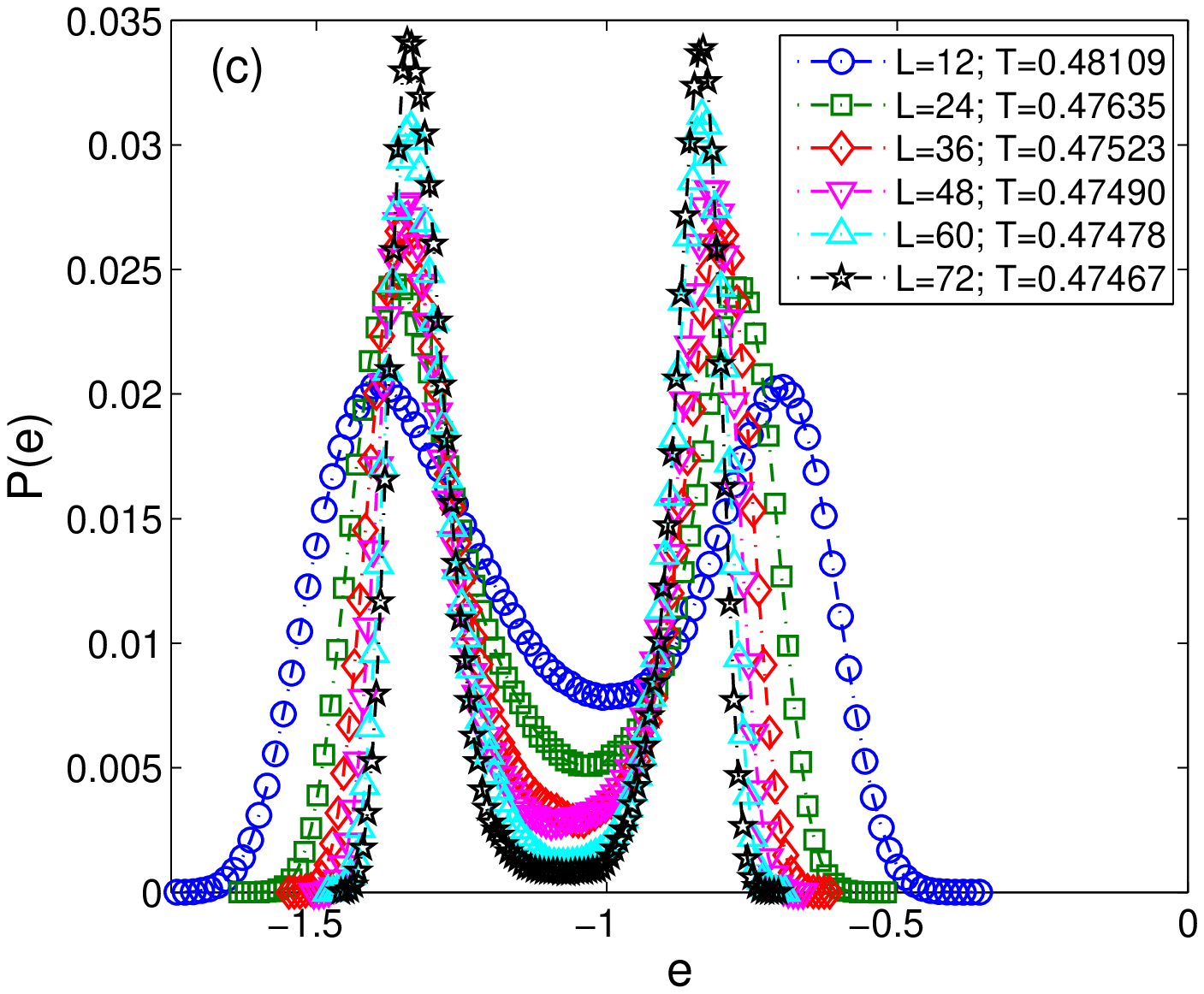}\label{fig:hist_alp_1_01_m100a}}
\subfigure{\includegraphics[scale=0.52,clip]{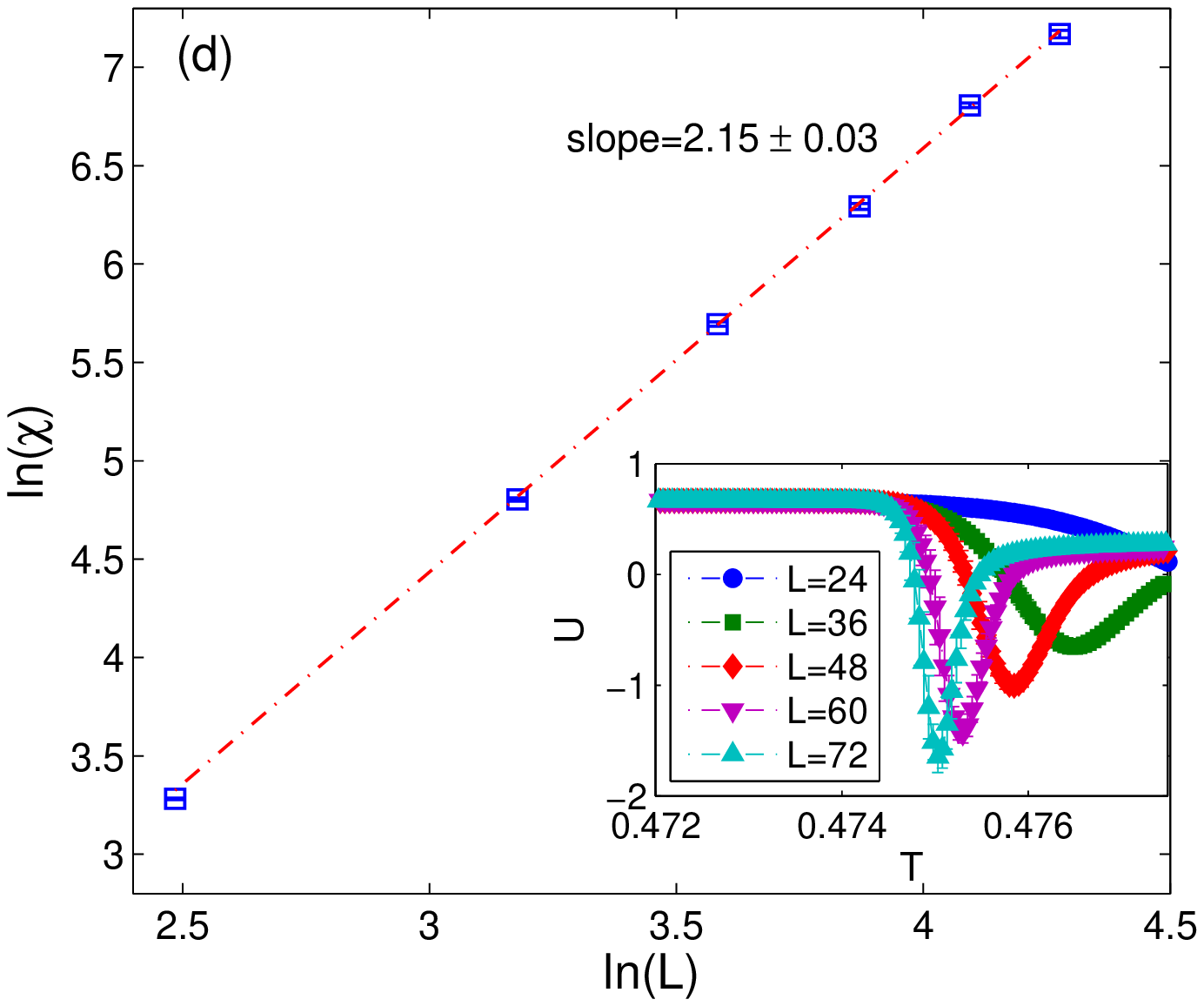}\label{fig:fss_chi_alp1_01_m100}}
\caption{(Color online) Energy histograms and FSS analysis for  $\alpha=1.01$ and (a,b) $p=50$ and (c,d) $p=100$. The histograms are reweighted to the temperatures at which the peaks are of equal height. The insets show the respective Binder cumulants.}\label{fig:hist_FSS_fin}
\end{figure} 

In Fig.~\ref{fig:hist_FSS_fin} we present the behavior at the transition for the cases of $p=50$ [Figs.~\ref{fig:hist_alp_1_01_m50a},~\ref{fig:fss_chi_alp1_01_m50}] and $p=100$ [Figs.~\ref{fig:hist_alp_1_01_m100a},~\ref{fig:fss_chi_alp1_01_m100}], at the value of $\alpha=1.01$. The respective features are very similar to those observed in Fig.~\ref{fig:hist_FSS_inf}, for the infinite $p$ case with $\alpha=1.03$ and $\alpha=1.02$, respectively. Namely, for $\alpha=1.01$ and $p=50$, all the measured quantities point to the continuous transition belonging to the BKT universality class, while for $\alpha=1.01$ and $p=100$, the transition is clearly of the first order. Therefore, for $\alpha=1.01$ the crossover value can be very roughly estimated as $50 < p_c < 100$.

\section{Summary}
We employed spin-wave theory and Monte Carlo simulations to study effects of inclusion of higher-order nearest-neighbor pairwise interactions with an exponentially decreasing intensity, $J_k=\alpha^{-k}$, where $\alpha>1$ and $k=2,\ldots, p$, to the standard XY model. At low temperatures, the spin wave theory predicts a quasi-long-range order phase characterized by an algebraically decaying correlation function with the exponent $\eta^{\rm sw} = T/(2 \pi J^{\rm sw})$, where $J ^{\rm sw} = \alpha/(\alpha-1) - p/(\alpha^p -1)$.

At higher temperatures, we showed that, in spite of belonging to the same universality class as the standard XY model, the studied generalized model can display qualitatively different behaviors, depending on the parameters $p$ and $\alpha$ that control the degree of nonlinearity. In particular, for a relatively small number of the higher-order terms $p$ and relatively fast decay of $J_k$, the critical behavior is qualitatively similar to that of the XY model, i.e., the system shows the Berezinskii-Kosterlitz-Thouless transition to the paramagnetic phase. Nevertheless, for $\alpha \to 1$ and $p$ large enough (not necessarily infinite), i.e., the parameters values corresponding to a highly nonlinear shape of the potential well, the transition changes to the first order. We demonstrated that the change of the transition order can be related to the behavior of topological excitations (vortices). Namely, in the parameter region where the potential well becomes very narrow the formation of vortex pairs at low temperatures becomes suppressed which facilitates their abrupt, discontinuous increase at the transition point.

\begin{acknowledgments}
This work was supported by the Scientific Grant Agency of Ministry of Education of Slovak Republic (Grant No. 1/0331/15) and the scientific grants of Slovak Research
and Development Agency provided under contract No.~APVV-0132-11 and No.~APVV-14-0073.
\end{acknowledgments}

\end{document}